\documentclass[prl,twocolumn,showpacs,epsfig]{revtex4}
\usepackage{graphicx}

\usepackage{graphics}
\usepackage{wrapfig}
\usepackage{dcolumn}
\usepackage{amssymb}
\usepackage{bm}
\usepackage{epsfig}
\usepackage{psfrag}
\usepackage{color}
\usepackage[utf8x]{inputenc}
\usepackage{ucs}
\usepackage{amsmath}
\usepackage{amsfonts}
\usepackage{amssymb}
\usepackage{graphicx}
\usepackage{dcolumn}
\usepackage{bm}
\usepackage{latexsym}
\usepackage{hyperref}

\def\deg{^{\circ}}

\def\lsim{\:\raisebox{-0.5ex}{$\stackrel{\textstyle<}{\sim}$}\:}

\def\3dots{\:\raisebox{-0.5ex}{$\stackrel{\textstyle.}{:}$}\:}
\def\beq{\begin{equation}}
\def\eeq{\end{equation}}
\def\bea{\begin{eqnarray}}
\def\eea{\end{eqnarray}}
\def\deg{^{\circ}}

\begin{document}

\title{Sorting motile rods by activity}
\author{Nitin Kumar$^1$\footnote{Presently at: James Franck Institute, University of Chicago, Chicago, Illinois 60637, USA}, Harsh Soni$^{1,2}$\footnote{Current address: School of Engineering, Brown University, Providence 02912, USA}, Rahul Kumar Gupta$^2$, Sriram 
Ramaswamy$^{2,\ddag}$ and A.K.
Sood$^1$}
\affiliation{$^1$Department of Physics, Indian Institute of Science, Bangalore
560 012, India}
\affiliation{$^2$TIFR Centre for Interdisciplinary Sciences, Tata
Institute of Fundamental Research, 21 Brundavan Colony, Osman Sagar
Road, Narsingi, Hyderabad 500 075, India}
\altaffiliation{On leave from the Department of Physics, Indian Institute
of Science, Bangalore}
\date{\today}
\begin{abstract}
	
We show, through experiments and simulations, that geometrically polar granular rods, 
rendered active by the transduction of vertical vibration, undergo a collective 
trapping phase transition in the presence of a V-shaped obstacle when the 
opening angle drops below a threshold value. We propose a mechanism that 
accounts qualitatively for the transition, based on the cooperative reduction 
of angular noise with 
increasing area fraction. We exploit the sensitivity of trapping to the 
persistence of directed motion to sort particles based on the statistical 
properties of their activity.
	
\end{abstract}
\maketitle

The interplay of directed energy transduction and interparticle interaction 
gives rise to a host of dramatic self-organizing effects in collections of 
active particles \cite{SriramAnnRev,SriramRMP}. Although the active matter 
paradigm was formulated to describe the living state, it is frequently more 
practical to study this class of systems by creating faithful imitations 
\cite{VJScience,DJDurian,BauschNature,DogicNature,NatCom}. Here we work with 
fore-aft asymmetric metal rods, millimeters in length, confined in a quasi 2D 
geometry and rendered motile in the horizontal plane by vertical vibration. 
Such objects, which we shall refer to as active polar 
rods \cite{VJthesis,VJScience,NatCom,Sano} are now a standard test-bed for 
probing the collective \cite{NatCom} and single-particle \cite{NKPRL,NKPRE} statistical 
physics of self-driven matter. The dynamics of self-propelled particles is 
influenced by the shape of the 
confining boundary, often leading to clustering and trapping 
\cite{austin,kudrolli}. Such effects could have 
practical consequences in the industrial processing of granular material 
through their influence on transport through narrow 
channels. Experiments with 
self-propelled granular particles have shown that clustering at the 
periphery can be minimized through the use of scalloped boundaries whose 
curvature reinjects particles into the 
interior \cite{dauchotPRL,NatCom,bristlebot}. However, a closer exploration of 
the interaction of active particles with boundaries and obstacles is called 
for. 

\begin{figure*}[t]
	\begin{center}
		\includegraphics[width=0.96\textwidth]{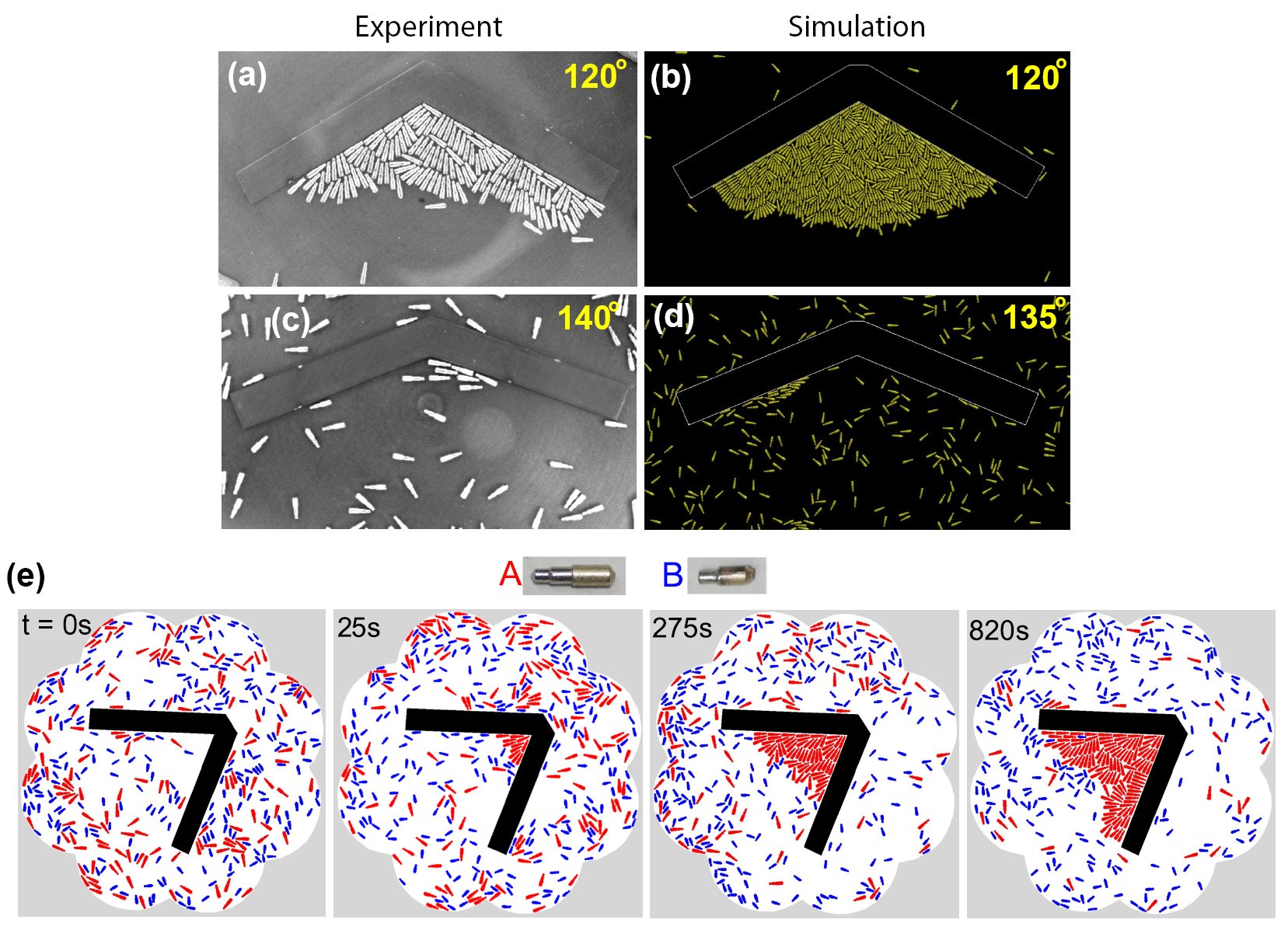}\\
		\caption{(a, b, c, d) A typical trapped and untrapped states in experiment and simulation. The angles are mentioned in yellow. The system size, in terms of rod length, is 20 and 39 in experiment and simulation respectively. (e) A sequence of images showing separation of polar particles based on their activity with only particle A getting trapped.}
		\label{results}
	\end{center}
\end{figure*}

In this Letter we investigate, experimentally and numerically, the collective 
dynamics of mono- and bidisperse collections of active polar rods in the 
presence of a V-shaped trap, as a function of the angle $\theta$ of the V 
and the nature of fluctuations in the motion of a single rod. In experiments the 
latter 
is governed by particle shape, while in simulations it can be varied 
continuously. Our results are as follows. (i) Particles with strongly 
directed motion are trapped for $\theta \lsim 120^{\circ}$, consistent with the 
active Brownian studies of \cite{capture} (Fig. \ref{results}, top panel). In particular, we offer
evidence to the effect that the onset of trapping has the character of a phase transition. (ii) Most 
interesting is the \textit{sorting} function of the trap: when 
placed in a homogeneous bidisperse mixture it rejects particles with noisy 
motion and collects those with persistent motility (Fig. \ref{results}, bottom panel). The idea that a 
passive enclosure can spontaneously sort \cite{potter} active particles based on the 
statistical characteristics of their motion suggests new directions for 
isolating motile cells or bacteria of different types. 

We now describe our findings in detail. Our experiments are carried out in 
a shallow circular geometry with 13 cm diameter \cite{NKPRL} and a flower 
shaped cell wall in order to avoid clustering \cite{NatCom}. It is covered by a 
glass lid at 1.2 mm above the surface, thus forming a confined two-dimensional 
system. We work with geometrically polar brass rods of length $\ell = 4.5$ mm 
and diameter 1.1 mm at the thick end. The cell is fixed on a permanent-magnet 
shaker (LDS V406-PA 100E) which drives the plate sinusoidally in the vertical 
direction with amplitude $a_{0}$ and frequency $f$= 200 Hz, corresponding to 
dimensionless shaking strength $\Gamma \equiv {a_{0}(2 \pi f)^{2}}/{g} = 7.0$ , 
where $g$ is the acceleration due to gravity. The rod transduces the vibration 
into predominantly forward motion in the direction of its pointed end 
(supplementary video 1).  A high-speed camera (Redlake
MotionPro X3) records the dynamics of the particles and ImageJ \cite{ImageJ} is used to extract instantaneous position, orientation and  velocity of the rods.

We use V-shaped traps of aluminium with arm length $L = 10 \ell \simeq 4.5$ cm 
with $20^{\circ} \leq \theta \leq 160^{\circ}$ in steps of $10^{\circ}$. 
The trap is placed in the middle of the cell and stuck to the surface with 
double-sided tape. The cell is filled by a layer of rods spread homogeneously 
and isotropically on the surface. All experiments on the trapping 
in monodisperse systems are done with the number of rods fixed at 150. 

\begin{figure}
	\begin{center}
		\includegraphics[width=0.48\textwidth]{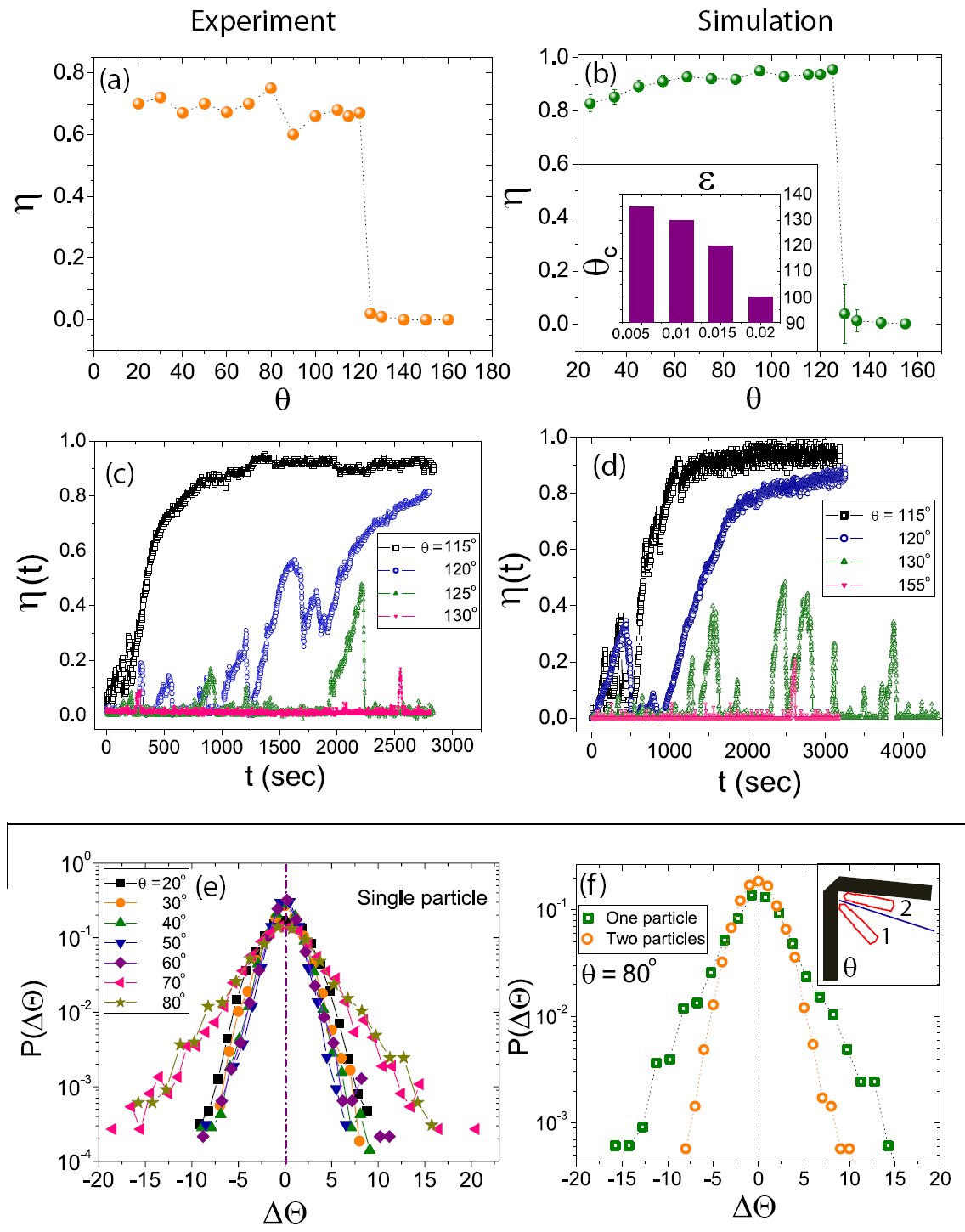}\\
		\caption{Trapping efficiency shows a sudden jump at $\theta = 
120\deg$ for (a) experiment and (b) simulation, indicating a trapping to 
detrapping
			transition. The critical transition angle decreases monotonically with angular noise as shown in the inset to (b). Close to the transition angle, we see repeated attempts where rods tend to form metastable structures
			inside the trap which become increasingly rare as we 
move away from the transition angle, for both experiment (c) and 
simulation (d). (e) Probability distribution of angular fluctuations of single 
rod 
			inside a trap. The transition angle of escape for one rod is $70\deg$ at and 
			beyond which the particle begins to show enhanced angular fluctuations. (f) 
			Introduction of second rod suppresses the width of the distribution which leads 
			to a trapped state. (Inset) A schematic of two rods in a trap.}
		\label{transition}
	\end{center}
\end{figure}

Mechanically faithful simulations, with details as in \cite{NatCom}, are 
conducted to investigate which properties of individual particles govern 
their propensity to get trapped and, hence, which features 
control activity-based sorting. 
We construct the tapered rods as arrays of overlapping spheres of 
different sizes \cite{NatCom}. Vibrating base and lid are represented by two 
hard horizontal walls whose vertical ($z$) positions at time $t$ are 
$\mathcal{A} \cos 2 \pi f t$ and $\mathcal{A} \cos 2\pi f t + w$ respectively. In our simulations we work at $\phi_r/F = 1.2$, where 
$\phi_r$ is rod area fraction and $F$ is the ratio of trap area $A_t = L^2 \sin 
\theta/2$ to that of the base, to ensure that the trap does not trivially 
exhaust the the total number of rods available on the base plate. We choose periodic boundary conditions 
in the $xy$ plane with linear dimensions of $19.3 \ell$, consistent with the 
experimental geometry. We also run simulations at larger system sizes, always at 
$\phi_r/F = 1.2$. Surface imperfections cause the rods in the experiments to 
perform rotational diffusion, which we capture in the 
simulation \cite{NatCom} by supplying a random angular velocity, 
$\omega_{z}=\varepsilon v_{rel}$ where $\varepsilon = 
\pm 0.0$1 mm$^{-1}$ with equal probability, whenever a rod collides with the base or the 
lid with relative velocity $v_{rel}$ of contact points normal to contact plane, 
and the value 0.01 is chosen to reproduce the observed orientational diffusion. 
We set the values of friction and 
restitution coefficients $\mu$ and $e$ to  0.05 and 0.3 for particle-particle 
collisions, 0.03 and 0.1 for rod-base collisions, 0.01 and 0.1 for rod-lid 
collisions and 0.03 and 0.65 for particle-V collisions 
respectively, to match the experiments as best we can. The ballistic dynamics 
of the particles is governed by Newtonian rigid body dynamics. VMD software 
\cite{VMD} is used to make all movies and snapshots from simulations.


Fig. \ref{results}(a) and (c) show the configuration of rods below and above the 
critical angle in experiments (See supplementary video 2).  Fig. \ref{results}(b) and (d) 
show the corresponding picture in simulation, with system size of $39\ell$ 
(supplementary video 3).

In order to quantify trapping, we calculate the instantaneous speeds of all the 
particles in every frame and track the number $N_0$ at zero velocity. We 
plot the trapping efficiency $\eta \equiv N_0 a / A_t$, where 
$a$ is the area of the two-dimensional projection of the rod on the 
surface. We find that 
this quantity drops abruptly at $\theta = 120\deg$ in experiment and 
$\theta = 125\deg$ in the simulation (Fig \ref{transition} (a) and (b) 
respectively). We study the effect of angular noise on the threshold 
angle $\theta_c$ below which trapping occurs. We find a monotonic decrease of 
$\theta_c$ with increasing angular noise $\varepsilon$; see inset to Fig. 
\ref{transition}(b). Beyond $\varepsilon = 0.02$ mm$^{-1}$ it is hard to observe 
trapping at any $\theta$. 

We have run longer experiments (45 min) and simulations at trap angles 
$115\deg, 120\deg, 125\deg$ and $130\deg$. We plot the time-evolution of $\eta$ 
in experiment Fig. \ref{transition}(c) and simulation Fig. \ref{transition}(d). 
For both experiment and simulation, with $\theta = 115\deg$, $\eta$ 
increases monotonically with time and saturates to a value 
close to 0.9. At $120\deg$ in the experiment the system displays multiple 
attempts, one of which leads to trapping. At somewhat larger 
$\theta$ in simulation and experiment such peaks in $\eta(t)$ are seen but are 
ultimately unsuccessful. These nucleation-like events together with the abrupt 
onset as a function of $\theta$ are indications of an underlying discontinuous 
nonequilibrium trapping phase transition. 

\begin{figure} [!b]
\begin{center}
\includegraphics[width=0.5\textwidth]{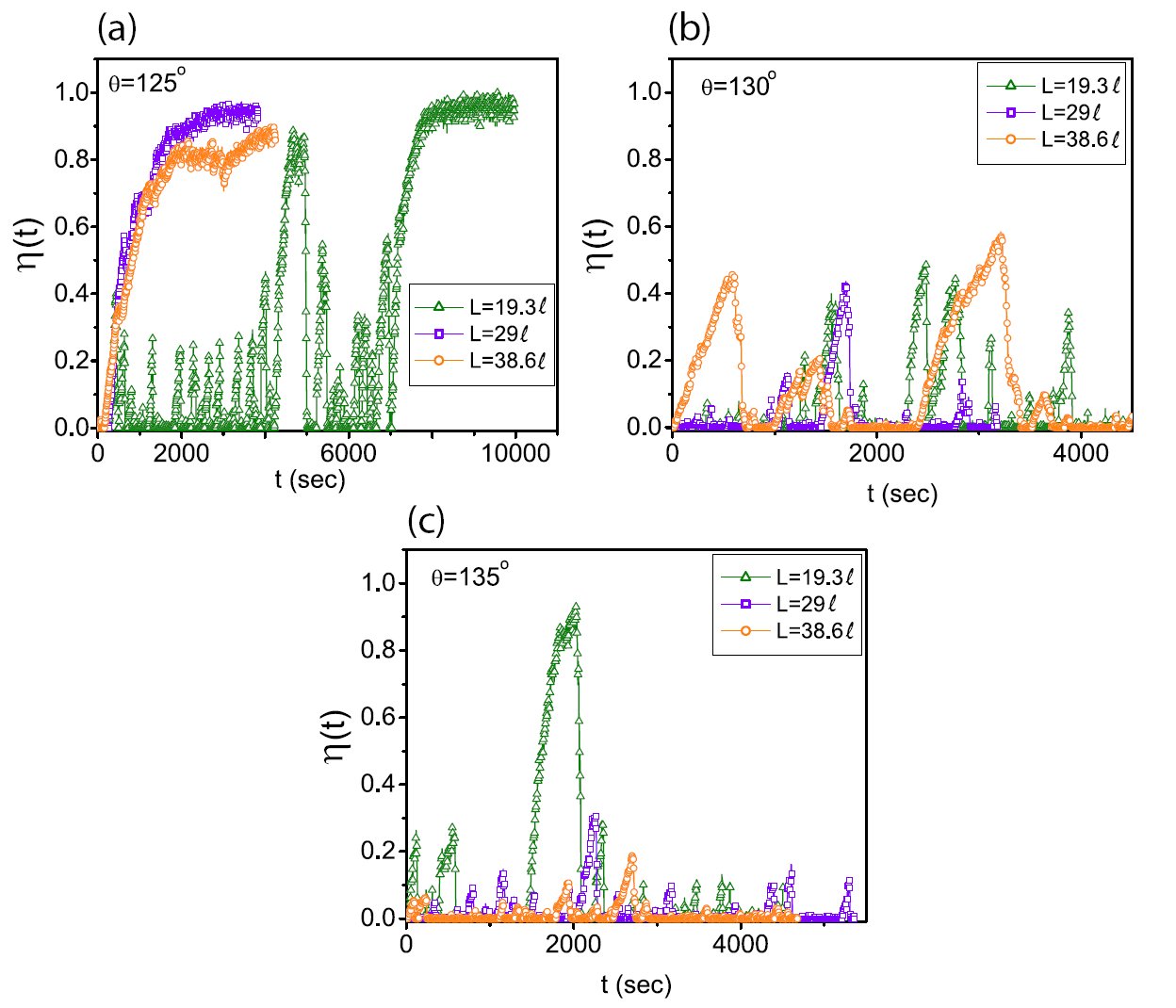}\\
\caption{(a) 
In a trapped state at angle $\theta=125\deg$. The value of $\eta(t)$ saturates 
to a constant value faster for bigger systems.  (b) In untrapped state near the 
phase boundary (at $\theta=130\deg$). The value of $\eta(t)$ fluctuates above 
zero but the height and the duration of spikes increases with systems size. (c) 
In untrapped state at $\theta=135\deg$. The height of spikes decreases with 
systems size. }
\label{systemsize}
\end{center}
\end{figure}

We turn now to the relation between single-particle and collective trapping. 
In Fig. \ref{transition}(e) we 
plot the probability distribution of angular fluctuations between two frames, 
i.e. P($\Delta\Theta$) where $\Delta\Theta(t)$ is the difference between the 
orientation of the rod in two successive frames (separated by 0.07 s). We find 
that for $\theta = 70\deg$ and $80\deg$, the width of the distributions is 
much larger than at smaller angles. We see that a single rod 
is trapped for $\theta \leq 70\deg$ (see supplementary video 4) for the 
duration of the experiment \cite{foot1}, and escapes for larger $\theta$ (supplementary video 5). 

Still at $\theta = 80 \deg$, we introduce a second rod [see inset to Fig. 
\ref{transition}(f)]. Interestingly, the particles cooperatively trap each 
other, each suppressing the angular fluctuations of the other [Fig. 
\ref{transition}(f) and supplementary video 6]. Physically, the effective trap angle 
seen by each rod is less than the imposed $\theta$, thereby 
preventing escape, as captured by the difference in the probability 
distribution $P(\Delta\Theta)$ for two cases in Fig. \ref{transition}(f). 
A na\"{\i}ve extrapolation suggests that as the number of particles is increased, a 
substantially larger $\theta$ should suffice to collectively trap a nonzero 
fraction of the rods. We do not however have a calculation that says that this 
threshold $\theta_c$ saturates around $120\deg$.  

To explore system-size dependence, of relevance to the question of whether 
the phenomenon is indeed a phase transition, we consider trap arm lengths 
$L=19.3\ell, 29\ell$ and $38.6\ell$, proportionately scaling base area and 
number of rods. 
The simulations were 
run for three values, below ($\theta = 125\deg$), above ($135\deg$) and 
close to the transition ($130\deg$). From the plot of $\eta$ as a function of 
time in Fig. \ref{systemsize} we see many failed nucleation attempts for 
$L = 19.3 \ell$, whereas for $L/\ell = 29$ and $38.6$ the trapping order 
parameter rapidly reaches a robust steady state value with 
fluctuations suppressed. This suggests a phase transition in the large$-L$ 
limit. 

It was remarked in ref. \cite{capture} that increasing angular noise favours escape 
from the trap. Such noise could enter in the form of run-and-tumble 
behaviour, conventional angular diffusion and/or translational diffusion. In our 
experiments, the noise depends in 
detail on particle shape. Given the sensitivity of trapping to these 
differences in particle properties, we ask whether a trap can sort particles 
based on their activity.  
\begin{figure}[!t]
\begin{center}
\includegraphics[width=0.5\textwidth]{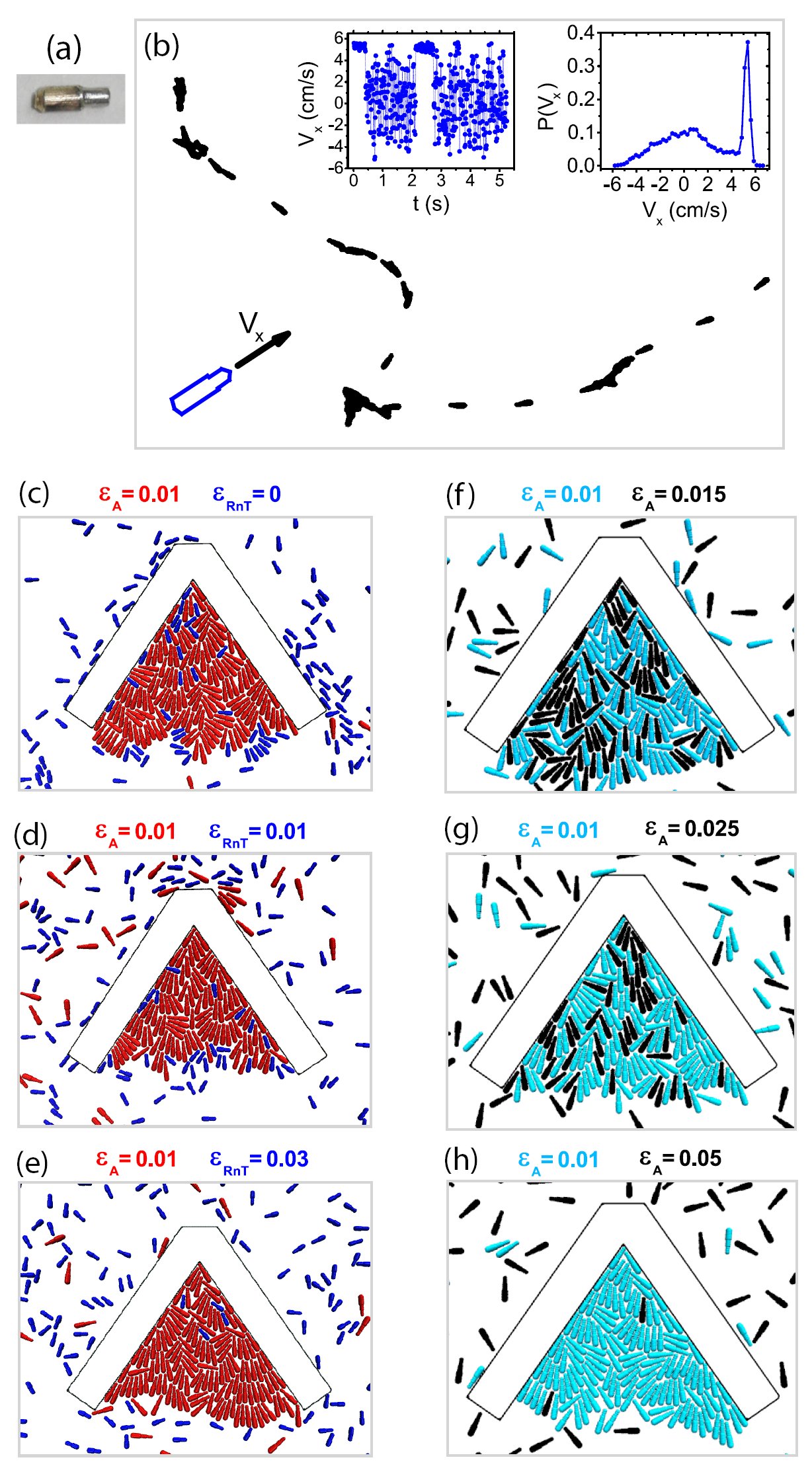}\\
\caption{(a) A photograph of the RnT particle along with its x-component of the in-plane velocity $V_x$ and orientation $\theta$. (b) A typical trajectory of the particle showing run and tumble events with time. Plot of $V_x$ as function of time and its probability distribution in the inset. (c,d,e) Snapshots of steady states of mixture of A and RnT for constant $\epsilon_A$ and different $\epsilon_{RnT}$. (f,g,h) Typical steady states for the case when angular noise is imposed selectively on particles A.}
\label{segg}
\end{center}
\end{figure}
We therefore introduce another active polar rod [see Fig. \ref{segg} (a)]. This 
particle is 3.5 mm long and 1.1 mm thick at its thicker end, and tapered in a 
single step, and displays dynamics qualitatively different from that 
of the particles (hearafter type A) discussed in the first part of this paper. 
Fig. \ref{segg} (b) shows a typical trajectory of this particle (see 
supplementary video 7), displaying rapid directed \textit{runs} interrupted by 
abrupt \textit{tumbles} following which a new run direction is selected at 
random. In what follows we will refer to these as RnT particles although, 
unlike in bacteria \cite{ecoli1,ecoli2}, the tumbles here are 
generally longer than the runs. The insets to Fig. \ref{segg}(b) show a 
typical time-trace of $V_{x}$, the instantaneous velocity component along the 
long axis, and the probability distribution $P(V_x)$. Note the strongly bimodal 
character of $P(V_x)$, with a sharp peak corresponding to the run motion. 
Experiments similar to those discussed above show that our RnT particles are 
not trapped for any $\theta$. 

We now introduce an initially homogeneous mixture of 150 type A and 225 RnT 
particles into the sample cell containing a trap with $\theta = 70\deg$. 
Upon vertical shaking we find a strongly selective trapping of only the A 
particles, with RnT entering and leaving freely. The four images in time 
sequence in Fig. \ref{results}(e) illustrate this separation, and the 
supplementary video 8 shows the kinetics in detail. 

In order to study what aspect of shape or kinetics governs the relative 
susceptibility to trapping, we carry out simulations in which we have 
independent control over particle properties. We do not attempt to recreate the 
complex dynamics of the RnT particles. We retain the shapes of the A and RnT 
particles, but force them with angular white noise. We keep the A particle 
noise at levels consistent with the experiment, and vary the noise strength on 
the RnT particles. As seen in Fig. \ref{segg}(c), (d) and (e) (and supplementary video 9) for zero noise, intermediate noise and high noise respectively, the sorting is highly effective in all 
cases, with no perceptible effect due to the angular noise on the RnT.
This is presumably a consequence of the sensitivity to initial 
conditions of their deterministic dynamics, through the interplay of 
agitation, shape and interaction with the bounding surfaces. This 
lack of persistence is what saves the RnT particles from being trapped. This is 
reminiscent of the escape strategy of myxobacteria \cite{myxo} when near an 
obstacle. Lastly, we study mixtures of geometrically identical A particles, 
distinguished only by an imposed difference in their angular noise. Again, Fig. \ref{segg}(f), (g) and (h) (and supplementary video 10), in increasing order of difference in noise strengths, show highly effective sorting at large noise difference, with the 
trap predominantly populated by the less noisy, more persistent component. 

Thus, the trappability of a particle is linked to the persistence of its 
directed motion. Reducing this persistence, whether through angular noise or 
enhanced shuffling along the axis of the particle, facilitiates escape from the 
trap, and results in a preferential accumulation of persistent movers inside 
the trap. 

In summary, our experiments and simulations find a phase transition to a 
collectively trapped state when a V-shaped obstacle is introduced amidst a 
monolayer of artificially motile macroscopic rods, as the V is narrowed or the 
angular diffusion of the rods reduced. Particles with highly directional 
motion are preferentially trapped; a mixture of particles with different 
motility characteristics is thus spontaneously sorted, concentrating persistent 
movers inside the trap and noisy particles outside. We offer a qualitative 
understanding of these phenomena, but a detailed theory of trapping and sorting 
as possible nonequilibrium phase transitions remains to be formulated. 

For support, N.K. thanks the University Grants Commission
(UGC), India, H.S. thanks the Council of Scientific and
Industrial Research (CSIR), India, and A.K.S. and S.R.
acknowledge J C Bose Fellowships from the Department
of Science and Technology (DST), India.


\begin{thebibliography}{50}
\bibitem{foot1} Our findings are consistent with a trapping of a nonzero 
fraction of rods in the infinite-size limit, amounting to a nonequilibrium 
phase transition. Whether the trapping of a single active rod is a phase 
transition like that of the delta-function potential in three dimensions is 
worth further study.
\end{thebibliography}
\end{document}